\documentclass[aps,amsfonts,prl,showpacs,nobibnotes,nofootinbib,%
tightenlines,twocolumn]{revtex4}
\input BoxedEPS
\SetRokickiEPSFSpecial  
\HideDisplacementBoxes
\begin{document}
\newcommand{\beq}{\begin{equation}}
\newcommand{\eeq}{\end{equation}}
\newcommand{\barr}{\begin{eqnarray}}
\newcommand{\earr}{\end{eqnarray}}
\def\figwidth{7.5cm}
\newcommand{\rjdel}[1]{[[   #1 ]]}
\newcommand{\rjcomm}[1]{{\sc   #1}}
\newcommand{\rjadd}[1]{{\bf #1}}
\newcommand{\ascomm}[1]{{\textsc{[[ AS: #1]]}}}
\newcommand{\asdrop}[1]{}
\newcommand{\asadd}[1]{[[#1]]}
 \newcommand{\andy}[1]{ }
\def\cH{{\cal H}}
\def\cG{{\cal G}}
\def\cV{{\cal V}}
\def\cU{{\cal U}}
\def\bra#1{\langle #1 |}
\def\ket#1{| #1 \rangle}
\def\coltwovector#1#2{\left({#1\atop#2}\right)}
\def\upp{\coltwovector10}
\def\downn{\coltwovector01}
\def\Ord{{\cal O}}
\def\bmp{\mbox{\boldmath $p$}}
\def\rhobar{\bar{\rho}}
\renewcommand{\Re}{{\rm Re}}
\renewcommand{\Im}{{\rm Im}}
\def\ask{\marginpar{?? ask:  \hfill}}
\def\fin{\marginpar{fill in ... \hfill}}
\def\note{\marginpar{note \hfill}}
\def\check{\marginpar{check \hfill}}
\def\discuss{\marginpar{discuss \hfill}}
\title{The Casimir Effect and Geometric Optics}
\author{R. L. Jaffe}
\author{A. Scardicchio} 
\email{jaffe@mit.edu, scardicc@mit.edu}
\affiliation{Center for Theoretical Physics \\ Massachusetts Institute
  of Technology \\ Cambridge, MA 02139, USA}
\begin{abstract}
	\noindent We propose a new approach to the Casimir effect based on
	classical ray optics.  We define and compute the contribution of
	classical optical paths to the Casimir force between rigid bodies. 
	We reproduce the standard result for parallel plates and agree
	over a wide range of parameters with a recent numerical treatment
	of the sphere and plate with Dirichlet boundary conditions.  Our
	approach improves upon the proximity force approximation.  It can be
	generalized easily to other geometries, other boundary conditions,
	to the computation of Casimir energy densities and to many other
	situations.
\end{abstract}
\vspace*{-\bigskipamount}
\preprint{MIT-CTP-3432}
\maketitle
\begin{figure}
\begin{center}
\BoxedEPSF{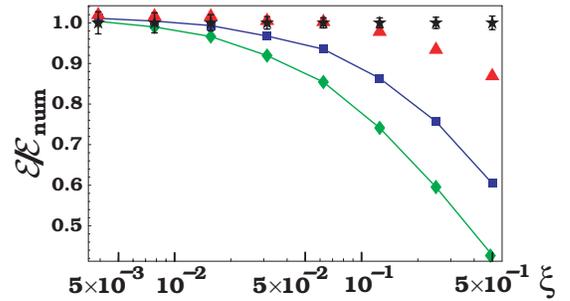 scaled 700} \caption{Casimir energy for a sphere
and a plate as a function of $\xi=a/R$.  Numerical data normalized to
unity (stars with error bars); Optical approximation (red triangles);
plate based PFA (green diamonds); sphere based PFA (blue squares) }
\label{fig:comparison}
\end{center}
\end{figure}

Improvements in experimental methods have rekindled efforts to compute
Casimir forces for geometries beyond the classic case of parallel
plates\cite{Casimir,expt1}.  No exact expressions are known even for
simple geometries such as two spheres or a sphere and a plane.  It is
therefore interesting to consider new ways of viewing the Casimir
effect and the approximation schemes that they motivate.  In this
Letter we present a new approach based on classical ray optics.  Our
approach avoids the infinities that have plagued Casimir 
calculations. 
Like ray optics it is most accurate at short wavelengths and where
diffraction is not important.  Our basic result, see eq.~(\ref{eq2}),
is simple and easy to implement.  It coincides with the well-known
proximity force approximation (PFA) \cite{Derjaguin} close to the
parallel plates limit.  Recently a precise numerical result has been
obtained for the Dirichlet Casimir energy of a sphere of radius $R$
separated from a plane by a distance $a$\cite{Gies03}.  This provides
us an opportunity to test our approximation.  The results are shown in
Fig.~(\ref{fig:comparison}).  They give us
encouragement that the optical approach may provide a useful tool for
estimating Casimir forces in situations where exact calculations are
not available.
We consider a scalar field of mass $m$ satisfying the wave equation,
$(-\nabla^{2}-k^{2})\phi(x)=0$, in a domain ${\cal D}\subset
\mathbb{R}^3$ bounded by disconnected surfaces, ${\cal S}_{1}$,
${\cal S}_{2}$, \ldots on which it obeys Dirichlet (or Neumann)
boundary conditions.  At the end we comment on the generalization to
conducting boundary conditions for the electromagnetic field.  The
Casimir energy can be written as an integral over $\delta\rho(k)$, the
difference between the density of states in ${\cal D}$ and the density
of states in vacuum.  This, in turn, can be related to an integral
over the Green's function\cite{scandurra}
\begin{equation}
         {\cal E}= \frac{1}{2}\hbar \int_{0}^{\infty}dk
        \omega(k)\frac{2k}{\pi}{\rm Im}\int_{\cal D} d^3x
         \ \widetilde G(x,x,k+i\epsilon)
         \label{eq1}
\end{equation}
where $\omega(k)=\sqrt{c^{2}k^{2}+m^{2}c^{4}/\hbar^{2}}$ and
$\widetilde G\equiv G-G_{0}$ is the difference between the Green's
function and the vacuum Green's function.  We work in three dimensions
although the generalization of our results to other dimensions is
straightforward.  
We look for an approximate solution of the wave equation that becomes
exact for infinite, planar surfaces.  We approximate the full
propagator as a sum of terms ascribable to different optical paths,
and satisfying the wave equation with an error of
$\Ord\left(1/(kR)^2\right)$ \cite{Born,Kline} where $R$ is a typical
curvature of the surface (e.g. the radius of the sphere in the
sphere$+$plane situation).  We are also neglecting diffractive
contributions, arising from the presence of sharp boundaries, and we
assume the absence of caustics in the integration domain.  We expect
this to give an excellent approximation for the Casimir energy when
$\kappa R$ is large (here $\kappa$ is the dominant wave number in the
Casimir energy integral, $\kappa\sim 1/a$ where $a$ is the minimum
distance between the surfaces).
 
\begin{figure}
\begin{center}
\BoxedEPSF{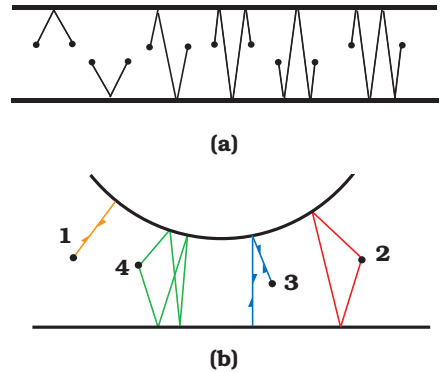 scaled 600} 
\caption{a) Optical paths for parallel plates.  Initial and final 
points have been separated for visibility; b) Optical paths for 
plane $+$ sphere.  For $n=1$ a reflection off the sphere is shown; 
for 
$n=3$ a reflection twice off the sphere is shown.}
\label{paths}
\end{center}
\end{figure}

The optical contribution to $G(x',x,k)$ is given by the sum over
\emph{optical} paths from $x$ to $x'$, $G(x',x,k) \to G_{\rm
optical}(x',x,k) = \sum_{\bf n}G_{\bf n}(x',x,k)$.  These fall into
classes, ${\cal C}_{n}$ which have $n$ points on the boundaries
\cite{Keller}.  The collective index ${\bf n}=(n,\alpha)$ identifies
both the class ${\cal C}_n$ and the path $\alpha$.  These paths are
stationary points in the class ${\cal C}_n$ of the functional integral
representation of $G$.  In three dimensions the optical terms in $G$
contribute,
\begin{equation}
G_{\rm optical}(x',x,k)=\frac{1}{4\pi}\sum_{\bf n} (-1)^{n}
\sqrt{\Delta_{\bf n}(x',x)} e^{ik\ell_{\bf n}(x',x)},
\label{eq4}
\end{equation}
(in $d\neq 3$ Hankel functions will appear in the analagous
expression\cite{Berry77}) where $\ell_{\bf n}(x',x)$ is the length of
the optical path ${\bf n}=(n,\alpha)$ that starts from $x$ and arrives
at $x'$ after reflecting $n$ times from the boundary.  These paths are
the minima of $\ell_{\bf n}(x',x)$, straight lines that reflect with
equal angles of incidence and reflection from the surfaces.  The
factor $(-1)^{n}$ implements the Dirichlet boundary condition.  For
Neumann boundary conditions it is absent.  $\Delta_{\bf n}(x',x)$ is
the enlargement factor of classical ray optics\cite{Kline} (also equal
to the VanVleck determinant as defined in Ref.~\cite{Berry77}), and is
given by
\beq
\Delta_{\bf n}(x',x)= \frac{d\Omega_{x}}{dA_{x'}}=\lim_{\delta\to 
0}\delta^{-2} e^{-\int_\delta^\ell 
ds\left(\frac{1}{R_1}+\frac{1}{R_2}\right)},
\label{enlarge}
\eeq
where  $R_{1,2}(s)$  are the radii of curvature of the wavefront 
following the path and $s$ is the coordinate along the path. 
$\Delta$ measures the spread in area $dA$ at the arrival point $x'$ 
of a
pencil of rays having angular width $d\Omega$ at the starting point
$x$, following the classical path indexed by ${\bf n}$.  This is
reasonably easy to compute even for multiple specular
reflections on any curved surface.
The contribution of the optical path ${\bf n}$ to the Casimir
energy is obtained by substituting eq.~(\ref{eq4}) into
eq.~(\ref{eq1}),
\begin{equation}
{\cal E}_{\bf n}=
 (-1)^{n}M_n \int_0^\infty \!
\!\!\frac{dk}{4\pi^{2}}\; \hbar k\omega(k)\!\!\!
 \int_{{\cal D}_{\bf n}} \!\!\!\! d^3x\!
\sqrt{\Delta_{\bf n}(x)} \sin k\ell_{\bf n}(x),
\label{eq5}
\end{equation}
where $M_n$ is the multiplicity of the $n$-th path, and we have
defined $\Delta_{\bf n}(x)=\Delta_{\bf n}(x,x)$ and $\ell_{\bf
n}(x)=\ell_{\bf n}(x,x)$ for brevity.  The minimum in the class ${\cal
C}_0$ (the direct path for $x$ to $x'$) should be excluded to account
for the subtraction of the vacuum energy.  In a given geometry the
optical paths can be indexed according to the number of reflections
from each surface.  For example in a geometry consisting in only two
convex plates (${\cal S}_{1}$ and ${\cal S}_{2}$) we have paths
reflecting once on ${\cal S}_{1}$ or once on ${\cal S}_{2}$, paths
reflecting two times (once on ${\cal S}_{1}$ \emph{and} once on ${\cal
S}_{2}$) and so on.  The multiplicity of even reflections paths is 2
(the path can be run in two different directions) while that of odd
reflections is 1.  ${\cal D}_{\bf n}$ is the domain over which the
path $\bf{n}=(n,\alpha)$ is possible.
${\cal E}_{\bf n}$ given by eq.~(\ref{eq5}) diverges if paths of
arbitrary small length can occur.  For domains bounded by convex
plates $\ell_{\bf n}(x,x)\to 0$ can only occur for the first
reflection, $n=1$.  To regulate this divergence we separate the
initial and final points by a distance $\epsilon$, so that $\ell_{\bf
n}\geq \epsilon$.  This is equivalent to putting a cutoff on the
frequency at $k\sim 1/\epsilon$\cite{BalianBloch,Deutsch79}.  Because
it is confined in the first reflection, the divergence will never
contribute to the force between surfaces.  In practice it can be
isolated and discarded.  Next we interchange the integrals over $k$
and $x$ in eq.~(\ref{eq5}), and perform the $k$-integral,
\begin{equation}
 {\cal E}_{\bf n}= (-1)^{n+1}M_n\frac{m^{2}c^{3}}{4\pi^{2}\hbar} 
\int_{{\cal
 D}_{\bf n}} d^3x  \frac{\sqrt {\Delta_{\bf n}(x)}}{\ell_{\bf 
n}(x)}K_2(m c\ell_{\bf n}(x)/\hbar).
\label{eq7}
\end{equation}
For a massless scalar we let $m\to 0$ and we obtain our fundamental 
result, 
\begin{equation}
		{\cal E}_{\rm optical}=-\frac{\hbar c}{2\pi^{2}}\sum_{\bf
		n}(-1)^{n} M_{n}\int_{{\cal D}_{\bf
		n}}d^{3}x\frac{\sqrt{\Delta_{\bf n}(x)}}{\ell_{\bf
		n}^{3}(x)}
		\label{eq2}
\end{equation}
which expresses the optical approximation to the Casimir effect as a 
sum over geometric quantities alone.
Our method should not be confused with Gutzwiller's semiclassical
approximation to the density of states\cite{Gutz,SandS}, nor with
Balian's and Bloch's multiple reflection expansion for the Green's
function\cite{BalianBloch}.  The latter expresses the Green's function
in terms of surface integrals, not limited to the optical paths.  The
former corresponds to performing the integration over $x$ in
eq.~(\ref{eq5}) by stationary phase, and selects only periodic paths. 
This approximation fails badly when the radius of curvature $R$ of the
surface(s) is large compared both to the separation $a$ and the width
$L$ of the surfaces.  Our method applies also in situations in which
no periodic classical paths exist.
Parallel plates provide a simple, pedagogical example which has many
features --- fast convergence, trivial isolation of divergences,
dominance of the even reflections --- that occur in all the geometries
we have analyzed.  We assume for simplicity that the two plates have the
same area $S$.  The relevant paths are shown in Fig.~(\ref{paths})
where the points $x$ and $x'$, which should coincide, are separated
for ease of viewing.  For the even paths $\ell_{2n}(z)=2na$,
$n=1,2,\ldots$, independent of $z$ (here $z$ is the distance from the
lower surface).  For the odd paths $\ell_{2n-1,\alpha}(z)=
2(n-1)a+2\zeta$, where $\zeta=z,a-z$ respectively if
$\alpha=\mathrm{down,up}$, $n=1,2,\ldots$ For planar boundaries the
enlargement factor is given by $\Delta_{\bf n}= 1/\ell^2_{\bf n}$.
The sum over even reflections,
\begin{equation}
{\cal E}_{\rm even}=-\frac{\hbar 
c}{\pi^{2}}\sum_{n=1}^{\infty}\int_{{\cal S}_d}
dS\int_{0}^{a}dz \frac{1}{(2na)^{4}}=-\frac{\pi^{2}\hbar c}{1440
a^{3}}S
\label{evencontr}
\end{equation}
is trivial because it is independent of $z$.  The result is the usual
Dirichlet Casimir energy\cite{expt1}.  The sum over odd reflections
gives
\begin{eqnarray}
{\cal E}_{\rm{odd}}&=&\frac{\hbar c}{2 \pi^2}\int
dS\sum_{n=0}^\infty
\int_{0}^{a}dz\frac{1}{\left(\epsilon^2+(2z+2na)^2\right)^2} 
\nonumber\\
&=&\frac{\hbar c}{16\pi^2}\frac{2\pi S}{\epsilon^3}.
\end{eqnarray}
The divergence as $\epsilon\to 0$ can be related to that expected on
the basis of Ref. \cite{BalianBloch} and will be discussed in
\cite{papertwo}. For now it suffices that it is proportional to
$S$, independent of $a$, does not contribute to the Casimir force, and
can be ignored.  The fact that the odd reflections sum to a divergent
constant is universal for geometries with planar boundaries, and to a
good approximation is also valid for curved boundaries.  Notice that
in this situation our method coincides with the method of images
\cite{Brown69}.  This will not occur for other examples.\footnote{The
same calculation was presented as an application of Gutzwiller's trace
formula \cite{Gutz} to the Casimir effect in Ref.~\cite{SandS}.  In
general the optical paths are closed but not periodic, and our
approach will not resemble Ref.~\cite{SandS} in other geometries.}
 
\begin{figure}
\begin{center} 
\BoxedEPSF{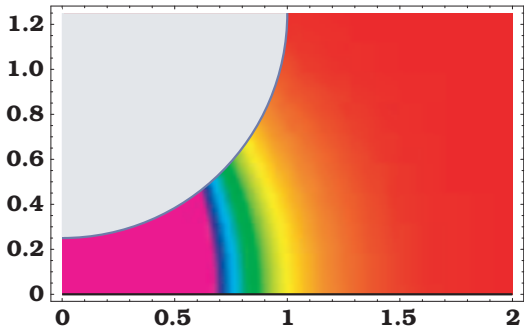 scaled 850}
\caption{Local contributions to the optical Casimir energy for a 
plane 
and a sphere with $a/R=0.25$.  The scale 
is linear in the hue from red (least) to violet (greatest). The
self-energy density given by ${\cal C}_1$ have been subtracted.}
\label{local}
\end{center}
\end{figure}

Note that the sum over $n$, eq.  (\ref{evencontr}), converges rapidly:
92\% of the effect comes from the first term (the two reflection path)
and $>98\%$ comes from the two and four reflection paths.  This rapid
convergence persists for all the geometries we an analyzed due to the
rapid increase in the length of the paths.  Also notice that for $m>0$
the two reflection contribution gives a uniform approximation to
the $a$ dependent part of the Casimir energy (accurate from $92\% $
for $ma\ll 1$ to exponentially small terms for $ma\gg 1$): ${\cal
E}|_{\rm{finite}}\simeq {\cal E}_2=-\frac{m^2 c^3}{8a^{2} \pi^2\hbar}
K_2 ( {2m ca}/{\hbar})$.
The optical approach sheds light on the proximity force approximation,
which has been used for years to estimate Casimir forces for
geometries in which an exact calculation is
unavailable\cite{Derjaguin}.  For the Dirichlet problem and two bodies
${\cal S}_{1}$ and ${\cal S}_{2}$ it takes the form:
\begin{equation}
        {\cal E}_{\rm PFA}(S_{1})=-\frac{\pi^{2}\hbar
        c}{1440}\int_{S_{1}}dS\frac{1}{[d_{12} (x)]^{3}}
\label{eq8}
\end{equation}
where $d_{12}(x)$ is the distance from ${\cal S}_{1}$ to ${\cal
S}_{2}$ along the normal to ${\cal S}_{1}$ at $x$.  The PFA is
ambiguous because a different result is obtained by interchanging
surfaces ${\cal S}_{1}$ and ${\cal S}_{2}$.  The PFA can be viewed as
the sum over optical paths if at each point $x$ in ${\cal D}$ the path
is chosen normal to ${\cal S}_{1}$ and the small area element that
intersects this path on ${\cal S}_{2}$ is replaced by a plane normal
to the path.  Then all paths that bounce back and forth between these
two parallel surfaces are summed.  Clearly the PFA misses three
important effects that are correctly included in the optical approach
a) the actual optical paths are shorter; b) the surfaces are curved;
and c) there are optical paths through points that do not lie on
straight lines normal to one surface or the other.  Effects a) and c)
increase and b) decreases for convex (increases for concave) surfaces
the optical estimate of the absolute value of the Casimir energy
relative to the PFA. In the cases we have studied the net effect is to
increase the Casimir energy.  In the subsequent discussions we compare
our results with the PFA approximations based on either of the two
surfaces.  First, however, we propose an ``optimal'' PFA motivated by
the optical approach: at each point in ${\cal D}$ choose the unique
\emph{shortest} path from ${\cal S}_{1}$ to ${\cal S}_{2}$ (of length
$\ell_{12}$).  Replace both surfaces locally by planes perpendicular
to this path and sum all optical contributions.  The result,
\begin{equation}
	{\cal E}_{\rm PFA^{*}}(S)=-\frac{\pi^{2}\hbar c}{1440}\int_{\cal D}
	d^{3}x \frac{1}{[\ell_{12}(x)]^{4}},
	\label{eq9}
\end{equation}
resolves the ambiguity in the PFA in favor of the shortest paths.  Of
course, the sum over the \emph{actual} optical paths including the
enlargement factor is more accurate still.
The only non-trivial geometry we know of for which we can test our
approximation is a sphere of radius $R$ placed at a distance $a$ from
an infinite plane.  This has not been solved analytically, but
numerical results have been published recently\cite{Gies03}.  Defining
$\xi=a/R$, we expect the optical approximation to give an error of
$\Ord\left(\xi^2\right)$. 
Certainly, when $\xi\gtrsim 1$ diffraction dominates, the force is given
by Casimir-Polder\cite{CP}, and our optical approximation will fail. 
Our aim is to study the accuracy of the optical approximation and the
domain of its applicability compared, for example, to the PFA.
We have calculated the Casimir energy for this configuration including
paths up to four reflections.  Some characteristic paths are shown in
Fig.~(\ref{paths}).  The results are plotted in
Fig.~(\ref{fig:comparison}).
The ${\cal C}_1$ and ${\cal C}_{3}$ contributions
can be evaluated analytically.  The divergent contribution from ${\cal
C}_{1}$ is independent of $a$ and can be put aside.  The
$a$-dependent, finite part of ${\cal E}_{1}$ and ${\cal E}_{3}$ are
opposite in sign and their sum is always small ($<2\%$) compared to
the ${\cal E}_{2}$.  ${\cal E}_{2}$ and ${\cal E}_{4}$ can be computed
quickly with Mathematica$^{\copyright}$.
 The optical result agrees with the numerical result of
Ref.~\cite{Gies03} within error bars ($\sim 1\%$) out to $\xi\approx
0.1$, where the PFA fails badly.  Even for $\xi\approx 1$, on the
border of its range of validity, $\delta E/E_{\rm N}=25\%$ where
$E_{\rm N}$ is the numerical value given by Ref.~\cite{Gies03} and
$\delta E=E_{\rm N}-E_{\rm opt}$.  In comparison, the ``sphere based''
PFA gives $\delta E/E_{\rm N}=58\%$ and the ``plate based'' PFA
$\delta E/E_{\rm N}=73\%$\footnote{In response to this paper, the
authors of Ref.~\cite{SandS} have compared their semiclassical
approximation with the numerical results of Ref.~\cite{Gies03} and
find agreement comparable to ours\cite{Sprivate}.  This can be
understood as a consequence of the fact that the effective width of
the sphere, $L$, scales like its radius $R$, and is not a general
feature of the semiclassical approach.}.
The limiting case $\xi\to 0$ is that of two infinite parallel plates. 
Here the optical approximation must agree analytically (and
numerically) with PFA. The agreement is clear in
Fig.~(\ref{fig:comparison}).
Since the optical approximation gives the Casimir energy as a volume
integral of a local contribution at each point $x$, it is possible to
get an idea of the domains that give the dominant contributions to the
Casimir force by plotting the integrand in eq.~(\ref{eq2}).  As an
example we show a contour map of the integrand for the plate and
sphere at $\xi=0.25$ in Fig.~(\ref{local}).  The local Casimir energy
density (and other local observables) is defined by a differential
operator acting on the Greens function\cite{Deutsch79} and is not
identical to the integrand in eq.~(\ref{eq2}).  However, since the
form of the optical approximation to the Greens function is so simple,
it is straightforward to obtain a compact and computable approximation
to the local energy density which would replace Fig.~(\ref{local}).
Our work suggests many possible extensions.  We have studied the
application to a finite (rectangular) plane inclined at an angle
$\theta$ to an infinite plane.  This is the geometry of a ``Casimir
torsion pendulum''\cite{papertwo}.
The physically interesting case of conducting boundary conditions can
be realized by constructing a matrix Green function for the vector
potential.  For parallel plates the contribution of odd reflections
integrates to $0$ (for any $a$), reflecting the well-known
absence of a $1/\epsilon^{3}$ divergence for conducting boundary
conditions\cite{Deutsch79}.  The even reflections sum to
obtain the expected result: ${\cal E}_{\rm cond}=2{\cal E}_{\rm
Dirichlet}$.  The extension to more complicated geometries will be 
presented in Ref.~\cite{papertwo}.
There are many interesting cases (both from a theoretical and an
experimental point of view) in which diffraction effects become
important.  Especially in situations in which the objects are small
compared to their separation (transition between Casimir and Van der
Waals forces).  Diffraction effects are certainly well beyond the PFA
approach but can in principle be included in the our framework by
using Keller's \cite{Keller} recipe for constructing the diffracted
rays contribution to the propagator  and then integrating  over $k$
to leave an $x$ integral.
The Casimir integral over modes can be generalized to give the
partition function for a fluctuating field at finite
temperature.  The computation is straightforward and gives the thermal
properties of the Casimir force --- assuming that it remains reasonable
to idealize the material by boundary conditions throughout the range
of temperatures of interest\cite{ackn}.
\acknowledgments
We thank P.~Facchi, S.~Fulling, I.~Klich, L.~Levitov, L.~S.~Schulman and
F.~Wilczek for discussions and suggestions.  We are grateful to
H.~Gies for correspondence and numerical values from the work of
Ref.~\cite{Gies03}, and to M.~Schaden for correspondence regarding 
Ref.~\cite{SandS}. This work is supported in part by the
U.S.~Department of Energy (D.O.E.) under cooperative research
agreements~\#DF-FC02-94ER40818.  A.~S. is a Bruno Rossi graduate
fellow supported in part by INFN.
 


\begin{thebibliography}{99}
	
\bibitem{Casimir} \andy{Casimir} H.B.G.Casimir,
  Proc. K. ned. Akad. Wet. {\bf 51}, 793 (1948).
\bibitem{expt1} S.\ K.\ Lamoreaux, Phys.\ Rev.\ Lett.  {\bf 78}, 5
(1997); For a review, see M.~Bordag, U.~Mohideen and
V.~M.~Mostepanenko, Phys.\ Rept.\ {\bf 353}, 1 (2001)
[arXiv:quant-ph/0106045].
\bibitem{Derjaguin} B.~V.~Derjaguin, Colloid Z.~{\bf 69} 155 (1934), 
B.~V.~Derjaguin,  I.~I.~Abriksova, and E.~M.~Lifshitz, Sov.~Phys.
JETP {\bf 3}, 819 (1957).
\bibitem{Gies03}
H.~Gies, K.~Langfeld and L.~Moyaerts,
JHEP {\bf 0306}, 018 (2003)
[arXiv:hep-th/0303264].
\bibitem{scandurra} See, for example, 
N.~Graham, R.~L.~Jaffe, V.~Khemani, M.~Quandt, M.~Scandurra and H.~Weigel,
Nucl.\ Phys.\ B {\bf 645}, 49 (2002)
[arXiv:hep-th/0207120].
\bibitem{Born} \andy{Born} M.~Born and E.~Wolf, \emph{Principles of
Optics}, Cambridge Univ.  Press.  (1980).
\bibitem{Kline} \andy{Kline} M.~Kline and I.~W.~Kay,
\emph{Electromagnetic theory and geometrical optics}, Interscience,
N.Y. (1965).
\bibitem{Keller} J.~B.~Keller, J. Opt. Soc. Am. {\bf 52}, 116 (1962);
J.B.Keller, in \emph{Calculus of Variations and its Application}
(Am. Math. Soc., Providence, 1958), p. 27; B.~R.~Levy and J.~B.~Keller,
Commun. Pure Appl. Math. {\bf XII}, 159 (1959); B.R. Levy and
J.~B.~Keller, Can. J. Phys. {\bf 38}, 128 (1960).
\bibitem{Berry77} M.~V.~Berry and K.~E.~Mount, Reps.~Prog.~Phys.  {\bf
35}
315 (1972).
\bibitem{BalianBloch} \andy{BalianBloch} R.~Balian and C.~Bloch, Ann. of
  Phys {\bf 69}, 401 (1970); {\bf 63}, 592 (1971); {\bf 69}, 76 
(1972).
  
  \bibitem{Deutsch79} \andy{Deutsch79} D.~Deutsch and P.~Candelas, Phys.
Rev.  D {\bf 20}, 3063 (1979).
\bibitem{Gutz}M.~C.~Gutzwiller, J. Math. Phys.{\bf 12}, 343 (1971);
\emph{Chaos in Classical and Quantum Mechanics} (Springer, Berlin,
1990).
\bibitem{Brown69} \andy{Brown69} L.~S.~Brown and G.~J.~Macay,
  Phys. Rev. {\bf 184}, 1272 (1969).
\bibitem{SandS} M.~Schaden and L.~Spruch, Phys. Rev. {\bf 58}, 935
  (1998); Phys. Rev. Lett. {\bf 84}, 459 (2000).
  \bibitem{Sprivate} M.~Schaden, private communication.
  
  \bibitem{CP}\andy{CP} H.~B.~G.~Casimir and D.~Polder, Phys. Rev. 
{\bf 73},
  360 (1948).
 
\bibitem{papertwo} R.~Jaffe and A.~Scardicchio, to be published.
\bibitem{ackn}  We thank I.~Klich and L.~Levitov for suggesting the 
last extension.
\end{thebibliography}
\end{document}